\documentclass[twocolumn,prl,superscriptaddress,showpacs,letter]{revtex4}

\usepackage{graphicx,nicefrac}
\usepackage{amssymb}
\usepackage{amsfonts}
\usepackage{xcolor}
\usepackage{amsmath}
\usepackage{siunitx}
\usepackage{soul}

\begin{document}

\title{High-resolution coherent probe spectroscopy of a polariton quantum fluid}

\author{F. Claude}
\affiliation{Laboratoire Kastler Brossel, Sorbonne Universit\'{e}, ENS-Universit\'{e} PSL, Coll\`{e}ge de France, CNRS, 4 place Jussieu, 75252 Paris Cedex 05, France}
\author{M. J. Jacquet}
\affiliation{Laboratoire Kastler Brossel, Sorbonne Universit\'{e}, ENS-Universit\'{e} PSL, Coll\`{e}ge de France, CNRS, 4 place Jussieu, 75252 Paris Cedex 05, France}
\author{R. Usciati} \affiliation{INO-CNR BEC Center and Dipartimento di Fisica, Università di Trento, via Sommarive 14, I-38050 Povo, Trento, Italy}
\author{I. Carusotto}\affiliation{INO-CNR BEC Center and Dipartimento di Fisica, Università di Trento, via Sommarive 14, I-38050 Povo, Trento, Italy}
\author{E. Giacobino}
\affiliation{Laboratoire Kastler Brossel, Sorbonne Universit\'{e}, ENS-Universit\'{e} PSL, Coll\`{e}ge de France, CNRS, 4 place Jussieu, 75252 Paris Cedex 05, France}
\author{A. Bramati}
\affiliation{Laboratoire Kastler Brossel, Sorbonne Universit\'{e}, ENS-Universit\'{e} PSL, Coll\`{e}ge de France, CNRS, 4 place Jussieu, 75252 Paris Cedex 05, France}
\author{Q. Glorieux}
\affiliation{Laboratoire Kastler Brossel, Sorbonne Universit\'{e}, ENS-Universit\'{e} PSL, Coll\`{e}ge de France, CNRS, 4 place Jussieu, 75252 Paris Cedex 05, France}

\begin{abstract}
Characterising elementary excitations in quantum fluids is essential to study collective effects within.
We present an original angle-resolved coherent probe spectroscopy technique to study the dispersion of these excitation modes in a fluid of polaritons under resonant pumping.
Thanks to the unprecedented spectral and spatial resolution, we observe directly the low-energy phononic behaviour and detect the negative-energy modes, i.e. the \textit{ghost branch}, of the dispersion relation.
In addition, we reveal narrow spectral features precursory of dynamical instabilities due to the intrinsic out-of-equilibrium nature of the system.
This technique provides the missing tool for the quantitative study of quantum hydrodynamics in polariton fluids.
\end{abstract}

\maketitle
The phenomenology of superfluids and Bose-Einstein condensates is generally understood in terms of quantum hydrodynamics.
From liquid helium to ultra-cold atoms, these hydrodynamic properties depend on the dispersion of the collective excitation modes in the quantum fluid, which, in the dilute regime, are described by the Bogoliubov theory~\cite{london_-phenomenon_1938,Landau1941,Bogolyubov:1947zz,feyn54}.
The Bogoliubov dispersion has a linear dependence on the wave-number ($k$) at low wave-number, where excitations are phonons (long-wavelength collective modes of the fluid), and a quadratic dependence at high wave-number, where excitations behave like free particles~\cite{Bogolyubov:1947zz,jin_collective_1996,mewes_collective_1996,onofrio_observation_2000,fontaine_disprel_2018,piekarski_measurement_2021}.
Quantum fluids may also be made of photons in nonlinear media, like exciton-polaritons that result from the strong coupling between cavity photons and excitons (bound electron-hole pairs in a semiconductor layer).
The direct detection of the Bogoliubov dispersion in polariton quantum fluids has remained a challenging goal ever since the observation of superfluidity~\cite{lagoudakis_quantized_2008,utsunomiya_observation_2008,amo_superfluidity_2009,kohnle_single_2011} and Bose Einstein condensation~\cite{kasprzak_boseeinstein_2006,balili_bose-einstein_2007} therein.

In this Letter, we present a high-resolution angle-resolved coherent probe spectroscopy technique that allows for the measurement of the Bogoliubov dispersion in the linear and nonlinear regimes of interactions.
In contrast to previous work~\cite{kohnle_four-wave_2012}, we continuously pump the polariton fluid near resonance to increase the spectral resolution; and unlike in photoluminescence experiments~\cite{stepanov_dispersion_2019,pieczarka_observation_2020,pieczarka_bogoliubov_2021}, we directly monitor the resonant response of the fluid to a coherent probe, thus efficiently isolating the signal on top of background emission from the fluid.
We obtain a dramatic increase in the spectral resolution and resolve the dispersion at all wave-numbers (down to $k=0$) and reveal the sound-like behaviour at low wave-number.
Furthermore, we show that the coherent probe may be used to generate Bogoliubov excitations on the so-called ghost branch of the fluid, therefore opening the way to a systematic study of the quantum depletion in driven-dissipative quantum fluids.
Finally, we observe additional narrow spectral features, elusive so far,  that go beyond the standard Bogoliubov theory of dilute gases~\cite{Bogolyubov:1947zz,pitaevskijBoseEinsteinCondensationSuperfluidity2016} and stem from the intrinsic non-equilibrium nature of polaritons~\cite{ciuti_quantum_2005,amelio_reservoir_2020}.

\paragraph*{Bogoliubov dispersion of collective excitations} --- 
At the mean-field level, the dynamics of polaritons in planar microcavities is governed by a Gross-Pitaevskii equation modified to account for their driven-dissipative nature --- the polariton fluid is a non-equilibrium system where the particle decay rate $\gamma$ is compensated for by continuous pumping with the laser field $E_p$ of frequency $\omega_p$ at normal incidence to the cavity plane ($k_p=0$).
Here, we focus on the lower polariton branch, well separated from the upper one by a polariton Rabi splitting of $\SI{5.1}{\milli\eV}$. 
The polariton density $n=|\psi|^2$ is related to the incident intensity $I=|E_p|^2$ by a nonlinear relation that depends on the detuning $\delta=\omega_p-\omega_{LP}$ of the pumping laser to the $k=0$ polariton mode~\cite{baas_bista_2004}: optical limiting is found for $\delta<0$ and optical bistability for $\delta>\sqrt{3}\gamma/2$.

The dispersion relation of Bogoliubov excitations in the fluid expresses the $k$-dependence of their frequency $\omega_B$ as
\begin{equation}
    \omega_B(k) = \pm\sqrt{\left(\frac{\hbar k^2}{2m^*} - \delta + 2 gn\right)^2 - (gn)^2} - i\gamma,
    \label{eq:Bog_disp}
\end{equation}
where $\hbar$ is the reduced Planck constant, $m^*$ is the polariton mass and $g$ is the polariton-polariton interaction constant~\cite{carusotto_quantum_2013}.
The real part of Eq.~\eqref{eq:Bog_disp}, the dispersion curve, has two branches $\omega_{B\pm}(k)$ symmetrical around the point $k=0$.
The Bogoliubov dispersion depends on the fluid properties through the density $n$ in Eq. ~\eqref{eq:Bog_disp}, so that the shape of the dispersion curve, as well as interaction regime, changes dramatically with detuning $\delta$.
This, in turn, will impact the propagation of collective excitations in the fluid~\cite{maitre_dark-soliton_2020,lerario_vortex-stream_2020,lerario_parallel_2020,jolyInterplayHawkingEffect2021}.

\begin{figure*}[ht!]
        \includegraphics[width=1\textwidth]{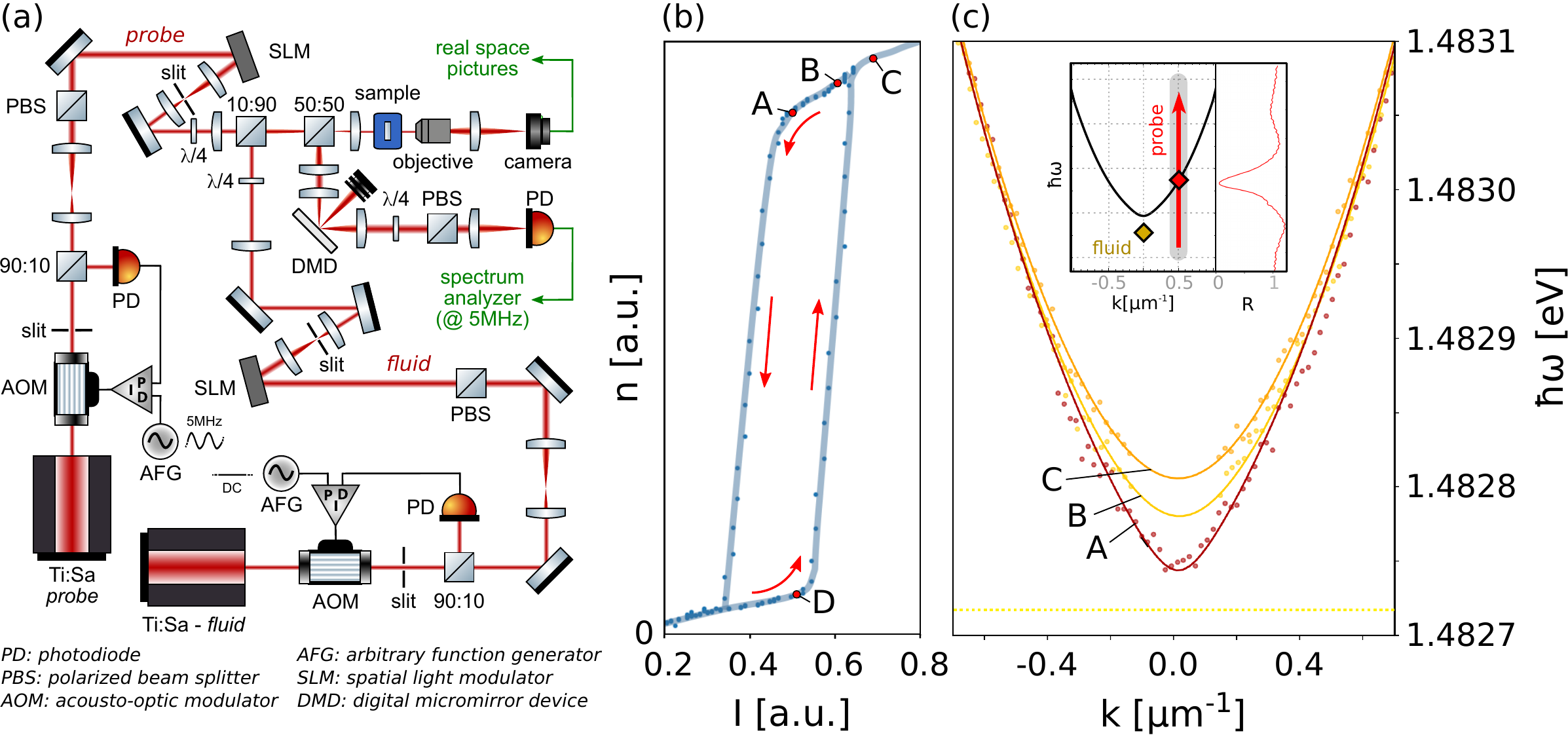}
        \caption{\textbf{Angle-resolved coherent probe spectroscopy}. \textbf{(a)} \textbf{Experimental setup}: The polariton fluid is pumped at normal incidence by a Ti:Sapph laser, whose spatial mode is controlled with a SLM. The fluid is probed by another Ti:Sapph laser whose intensity is modulated by an AOM and whose angle of incidence is controlled by a second SLM. Light in the reflection port is spatially filtered with a DMD and demodulated on an electronic spectrum analyzer. \textbf{(b)} \textbf{Optical bistability of the polariton fluid}. Polariton density as function of the pumping intensity. Red arrows indicate the hysteresis cycle direction. \textbf{(c)} \textbf{Bogoliubov dispersion} for a pump intensity $I$ giving a fluid density $n$ at working points A, B and C along the bistability cycle. Dispersion curves are constructed by collating responses to scans of the probe frequency and wave-number. Yellow dashed line, pump energy $\hbar\omega_p$. Insets: scan example and corresponding reflectivity spectrum; orange diamond: fluid laser frequency.
        }
    \label{fig:setup}
\end{figure*}

\paragraph*{High resolution spectroscopy ---}
The fluid is created by a resonant circularly polarised continuous-wave laser, and we use a second laser to stimulate the Bogoliubov elementary excitations.
This coherent probe is also continuous-wave and has the same polarisation as the fluid.
Its incidence angle $k_{pr}$ may be tuned at will and its  frequency $\omega_{pr}$ can be scanned continuously over a wide range while monitoring its reflectivity.
The response of the system is enhanced when the probe is resonant with a collective excitation $\omega_{pr}(k_{pr})=\omega_B(k_{pr})$, inducing a drop (peak) in the cavity's reflectivity (transmittivity).
The spectral position of the reflectivity dips as a function of $k_{pr}$ directly gives the Bogoliubov dispersion.
In terms of many-body theory,  the coherent probe beam couples to the polariton field component at $k_{pr}$ and the system response is read out on the polariton field, therefore this method provides a direct measurement of the impulse response (i.e. the Green's function) of the polariton field.

Experimentally, at the output of the cavity, the signal drops (peaks) are dwarfed by the intense bath of photons stemming from the fluid beam.
To counter this, we modulate the probe's amplitude
with an acousto-optic modulator (AOM) operating at a frequency $f_{\mathrm{mod}}$ significantly higher than the spectral width of the fluid laser.
Light coming out of the cavity is focused on a photodiode connected to an electronic spectrum analyser that demodulates at $f_{\mathrm{mod}}$, thus separating the signal from the background fluid.
Meanwhile, the angular resolution of the detection system is precisely controlled by means of a digital micromirror device (DMD) positioned in the Fourier plane of the cavity.

\paragraph*{Experimental implementation ---}
The setup is shown in Fig.~\ref{fig:setup} \textbf{(a)}.
The sample is a semiconductor microcavity made of two highly reflecting planar GaAs/AlGaAs Bragg mirrors, within which three InGaAs quantum wells separated by a substrate made of GaAs are placed at the three antinodes of the field.
The cavity finesse is on the order of 3000.
The polariton mass within is $m^*=6.0\times10^{-35}$kg.
The experiments are performed in a open-flow helium cryostat ($\mathrm{T}=4$K).
The polariton fluid is pumped with a collimated, continuous-wave Ti:Sapphire laser (linewidth less than $\SI{1}{\mega\hertz}$) that illuminates the cavity at normal incidence, creating a spot of diameter $\SI{100}{\micro\meter}$.
The transverse intensity profile of the fluid beam is controlled with a spatial light modulator (SLM), and operation is done with either a Gaussian or a top-hat mode.
The coherent probe is another continuous-wave Ti:Sapphire laser whose Gaussian spot has a $\SI{50}{\micro\meter}$ waist and whose amplitude is modulated by an AOM at $f_{\mathrm{mod}}=\SI{5}{\mega\hertz}$.
The frequency of the probe is continuously scanned over a range of $\SI{120}{\giga\hertz}$  ($\sim$ 0.5 meV) around the fluid's frequency $\omega_p$.
The angle of incidence of the probe field on the cavity is controlled with a second SLM.

Here we present reflectivity data.
As shown in Fig.~\ref{fig:setup} \textbf{(a)}, the reflected light is separated from incoming light by a 50/50 beam splitter placed before the cryostat.
In the detection arm, the DMD is used as a $k-$tunable circular pinhole of controllable position and diameter.
Placed in the reciprocal space of the cavity plane, it selects the reflected light at the wave-number $k$ on which the displayed pinhole is centred.
When the pinhole and the probe are at the same $k$, four-wave mixing emission that may appear at $-k$ is cut off.
As for the spectral resolution of the measurements, our technique does not rely on an optical spectrometer, and it is only limited by the spectral width of the probe laser, here narrower than 250 kHz ($< \SI{1}{\nano\eV}$), while the angular resolution is given by the diameter of pinholes displayed on the DMD, here $\delta k = 0.0189\pm\SI{0.0005}{\per\micro\meter}$.

\paragraph*{Shape of the dispersion curve} --- 
Since we create the polariton fluid with near-resonant pumping, we are able to study the shape of the dispersion curve as a function of the detuning $\delta$ and the pump intensity $I$.
In order to operate in the bistable regime, one must have $\omega_p>\omega_{LP}$, such that $\delta>\sqrt{3}\gamma/2$. 
We can reach the hysteresis region of the bistable loop by increasing  $I$ above the bistability threshold and then decreasing it back until the turning point.
This procedure must be repeated each time $\delta$ is modified.
Below, we first present the results for $\delta=\SI{0.2}{\milli\eV}$.

Fig.~\ref{fig:setup} \textbf{(b)} and \textbf{(c)} show the bistability curve of the polariton fluid and typical dispersion curves $\omega_{B+}(k)$ corresponding to working points A, B and C.
The insets in Fig.~\ref{fig:setup} \textbf{(c)} show a scan of the coherent probe frequency $\omega_{pr}$ at a given $k_{pr}$ and the response in the cavity's reflectivity --- when $\omega_{pr}(k_{pr})=\omega_B(k_{pr})$, a dip centred at the energy of the resonance is observed in the reflectivity.
The dispersion is reconstructed by collating these scans (separated by $\delta k=\SI{0.019}{\per\micro\meter}$) and tracking the minima of the dips at each $k_{pr}$, represented by the dots in Fig.~\ref{fig:setup} \textbf{(c)}.
Here the fluid is pumped with a Gaussian mode and the solid lines in Fig.~\ref{fig:setup} \textbf{(c)} are the moving average over 5 experimental points, $\omega_{B+}(k)$.

As expected from the theory~\cite{ciuti_quantum_2005}, there is a gap between the pump energy (horizontal yellow line) and the minimum of the $\omega_{B+}(k)$ branch at $k=0$.
Even though the size of the gap decreases when the pump intensity is decreased from working point C to A, it  is still present  at working point A near the turning point of the bistability.
Although the density distribution of the fluid is steepened by the bistable behaviour of the system, creating the polariton fluid with a Gaussian spatial mode does yield a non-uniform density~\cite{stepanov_dispersion_2019}, whence the dispersive properties of the fluid depend on space.
In other words, the dispersion curves shown in Fig.~\ref{fig:setup} \textbf{(c)} result from a spatial average over the different dispersion relation across the probe spot.
Even though, the shape of the dispersion at low $k,\omega$ is not parabolic, which is a hint of collective effects due to phononic interactions, the linear behaviour at low energy is only partially recovered and spatial resolution is required to probe the fluid with a fixed local density.
\begin{figure*}[ht!]
    \centering
    \includegraphics[width=\textwidth]{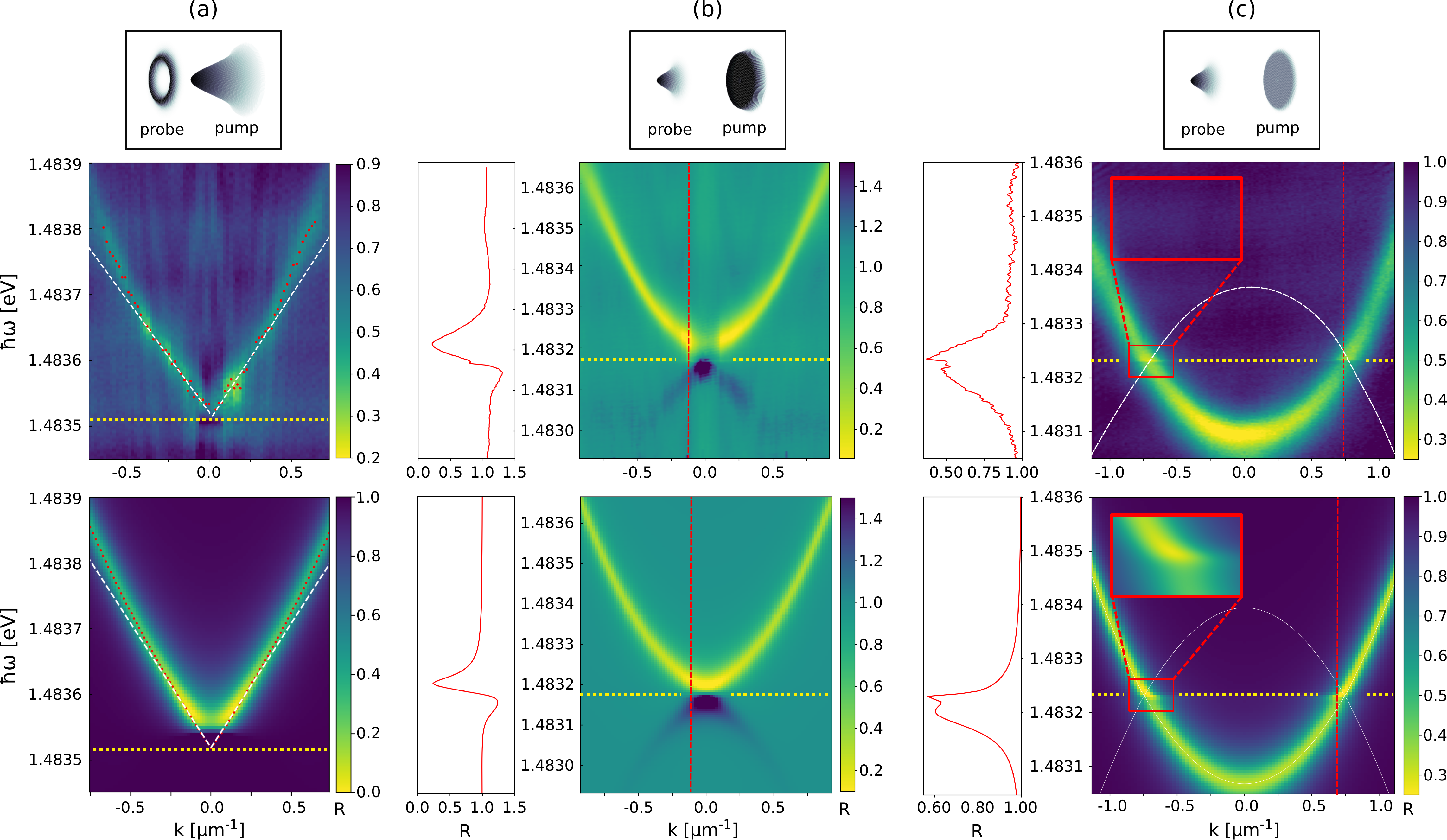}
    \caption{\textbf{Bogoliubov dispersion}. 
    Top row: spatial mode profile of the pump and coherent probe lasers. 
    Middle row: collated probe reflectivity scans separated by $\delta k=\SI{0.019}{\per\micro\meter}$.
    Bottom row: corresponding numerical simulations.
    Vertical red-dotted lines: $k$-slice corresponding to the energy scan shown on the left panel.
    Horizontal yellow dashed line: fluid pumping laser energy $\hbar\omega_p$. 
    \textbf{(a)} \textbf{Sonic dispersion}. Density at turning point A of the bistability loop, with $\delta=gn+g_r n_r$.
    White dotted line: linear dispersion fit of the speed of sound. 
    R is normalized to 1 for the off-resonance reflected probe intensity. 
    \textbf{(b)} \textbf{Ghost branch}. 
    Density near the turning point A of the bistability loop. 
    The ghost branch is coherently seeded by the probe and is visible at $\omega<\omega_p$. 
    \textbf{(c)} \textbf{Precursors of instabilities}. 
    Low density regime. 
    $k=\pm\SI{0.74}{\per\micro\meter}$ plateaus are visible at $\omega=\omega_p$. 
    Insets: band-sticking Fano features. 
    White dotted curve, position of the ghost branch.}
    \label{fig:dispcurves}
\end{figure*}

In order to observe the closing of the gap between $\omega_{B+}(k)$ and $\omega_p$, the region of spatial integration has to be reduced so as to measure the dispersion for a specific value of the local density $n$.
To this end, the probe intensity profile is reshaped with the SLM into a narrow ring.
The radius of the ring is chosen to match the outer edge of the fluid, where the local fluid density is as close as possible to the turning point of the bistability cycle ($\delta=gn$).
In Fig.~\ref{fig:dispcurves} \textbf{(a)}, we present the coherent probe reflectivity spectra in this configuration.
The small red dots show the position of the dip minimum for each frequency scan.
We observe the closing of the energy gap and we evidence directly, without any data processing, that $\omega_{B+}(k)$ is linear at low wave-number (white dashed line) and parabolic at large wave-number.
The agreement with numerical simulations based on a driven-dissipative version of the GPE~\cite{amelio_reservoir_2020} (bottom row of Fig.~\ref{fig:dispcurves} \textbf{(a)}) is excellent.

The speed of sound in the fluid is precisely obtained from the linear slope at low energy: $c_s=\SI{0.54}{\micro\meter\per\pico\second}$.
Interestingly, this value is reduced by a factor $\alpha$ with respect to the theoretical prediction $c_s^{th}=\sqrt{\delta/m_{LP}}$.
This can be explained in terms of an incoherent reservoir of long-lived dark excitons providing an additional contribution to the blue-shift and to the bistability curve but not to the speed of sound~\cite{stepanov_dispersion_2019,pieczarka_bogoliubov_2021,amelio_reservoir_2020}.
In our experiment, an estimate of the reservoir contribution to the blue-shift $g_rn_r$ = 0.75 $\delta$ is obtained from the measured values of $\alpha\simeq 0.49$ and of the total blue-shift $gn+g_rn_r=\delta$ at the  turning point of the bistability.

\paragraph*{Detection  of the ghost branch} --- The discussion so far has focused on $\omega_{B+}(k)$, the positive frequency solution of the dispersion relation~\eqref{eq:Bog_disp}, i.e. resonant excitations above the ground state.
Since we excite the system with a coherent probe coupled to the polariton field (and not with a time-dependent external potential as in standard Bragg spectroscopy of ultracold atomic gases~\cite{PhysRevLett.82.4569}), our method also enables the observation of the negative-energy solution $\omega_{B-}$  that is often referred to as the `ghost branch'.
Unlike the $\omega_{B+}$ branch, the ghost branch does not resonate with the cavity, and is spontaneously populated by quantum depletion of the fluid.
This renders its direct observation in photo-luminescence difficult~\cite{pieczarka_observation_2020,pieczarka_bogoliubov_2021,stepanov_dispersion_2019}.
Here, we stimulate the coherent conversion of the fluid of polaritons toward the ghost branch via the coherent probe~\cite{zajac_parametric_2015}:
this configuration induces a four-wave mixing (FWM) process ($\omega_p$, $\omega_p$)  $\rightarrow$ ($\omega_+$, $\omega_-$), and the probe at $\omega_-<\omega_p$ directly seeds Bogoliubov excitations on the ghost branch which manifest as an amplification peak $R>1$ above the baseline of the probe's reflectivity spectrum.

Within the Bogoliubov theory, this FWM process consists in the mixing of creation and destruction operators of the polariton field.
Due to the phase matching condition, this mixing is only sizeable within a limited wavevector region, so efficient angular selectivity is crucial.
In addition, by using a top-hat intensity profile we create a fluid of homogeneous density.  As a result, the FWM process takes place in a reduced range of interaction energies, thus optimising the population of the ghost branch and its detection.
In Fig.~\ref{fig:dispcurves} \textbf{(b)}, the fluid is pumped with an intensity as close as possible to the turning point of the bistability, where the FWM is most efficient.
Both branches of the dispersion relation at $\omega_{B\pm}$ are clearly visible, in good agreement with the numerical simulations shown in the bottom row of Fig.~\ref{fig:dispcurves} \textbf{(b)} and the strength of the ghost branch decreases rapidly with $k$, similarly to the behaviour observed in cold atoms~\cite{lopes2017quantum,cayla2020hanbury,tenart2021observation}.

\paragraph*{Fluid instabilities} --- Because of the driven-dissipative nature of polaritons, the quantum fluid is intrinsically out-of-equilibrium. 
This explains the gap that opens between the dispersion curve and the pump energy in the nonlinear regime of interactions (cf Fig.\ref{fig:setup} \textbf{(c)}), a marked departure from the standard Bogoliubov theory of dilute gases.

Even more interestingly, out-of-equilibrium physics is at the origin of dynamical instabilities of the fluid.
When operating in a regime of weak interactions ($\delta>3gn$)  corresponding to working point D in Fig.~\ref{fig:setup} \textbf{(b)}), the $\omega_{B\pm}$ branches cross around $k=\pm\SI{0.74}{\per\micro\meter}$ and $\omega=\omega_p$, and stick together for $\delta-3gn<\nicefrac{\hbar k^2}{2m^*}<\delta-gn$ ~\cite{ciuti_insta_2001,savvidis_insta_2001}, as in Fig.~\ref{fig:dispcurves} \textbf{(c)}.
There, the sign inside the square root in Eq.~\eqref{eq:Bog_disp} becomes negative: the real parts $\mathrm{Re}(\omega_B)$ both stay at $\omega=\omega_p$, yielding plateaus in the dispersion curve, while the imaginary parts $\mathrm{Im}(\omega_B)$ are split.
The negative imaginary-part root begets absorption, leading to the narrow Fano-like feature in the reflectivity (cf insets) while the positive imaginary-part root yields gain, hence the narrow peaks in the reflectivity.
This is an important marked departure from the physics of dilute gases.
Although, in our experiment in Fig.~\ref{fig:dispcurves} \textbf{(c)} the pump intensity is kept low enough for the fluid to remain dynamically stable, our high-resolution spectroscopy method reveals the observation of this out-of-equilibrium feature for the first time.
For slightly higher pump intensity, these plateaus would lead to dynamical instabilities predicted in~\cite{ciuti_quantum_2005} so our method is well suited for the in-depth analysis of this instability as well as other turbulent dynamics, such as snake-instabilities~\cite{claude_taming_2020} or quantum turbulence~\cite{madeira_quantum_2020,koniakhin_2d_2020,galantucci_new_2020,koniakhin_topological_2021,giacomelli_interplay_2021}.

\paragraph*{Conclusion} --- 
We have presented a angle-resolved coherent probe spectroscopy method to measure the dispersion of collective excitations in a quantum fluid of exciton-polaritons within a semiconductor microcavity.
The high energy- and wave-number-resolution of our spectroscopy method enables the accurate study of the Bogoliubov dispersion in all regimes of interactions, from the parabolic dispersion of high-energy single-particle excitations to the sonic dispersion of long-wavelength collective excitations, and to the negative-norm ghost branch resulting from the stimulated quantum depletion of the fluid.
We have observed the influence of the polariton fluid local density on the dispersion at low wave-number and evidenced the presence of a dark excitonic reservoir that modifies the effective speed of sound in the fluid.
The versatility of our method is further illustrated by the resolution of narrow spectral features that specifically stem from non-equilibrium effects such as the controlled opening of an energy gap and the appearance of horizontal plateaus precursory of dynamical instabilities.
Our method will be an essential tool in addressing the non-equilibrium physics of Goldstone modes induced by spontaneous phase symmetry-breaking~\cite{wouters_goldstone_2007} and for the study  of quantum amplification effects such as Hawking radiation or rotational superradiance in analogue gravity scenarios like (rotating) black holes~\cite{nguyen_acoustic_2015,jacquet_polariton_2020,jolyInterplayHawkingEffect2021}.

\begin{acknowledgements}
We acknowledge financial support from the H2020-FETFLAG-2018-2020 project ”PhoQuS” (n.820392). IC acknowledges financial support from the Provincia Autonoma di
Trento and from the Q@TN initiative. QG and AB are members of the Institut Universitaire de France.  
\end{acknowledgements}

\providecommand{\noopsort}[1]{}\providecommand{\singleletter}[1]{#1}%

\end{document}